# The influence of illumination conditions in the measurement of built-in electric field at p-n junctions by 4D-STEM

Bruno C. da Silva[1], Zahra S. Momtaz[1], Lucas Bruas[2], Jean-Luc Rouviere[3], Hanako Okuno[3], David Cooper[2] and Martien I. den-Hertog[1]

[1]*Univ. Grenoble Alpes, CNRS-Institut Néel, F-38000 Grenoble, France*
[2]*Univ. Grenoble Alpes, CEA-LETI, F-38000 Grenoble, France*
[3]*Univ. Grenoble Alpes, CEA, MEM, LEMMA F-38000 Grenoble, France*

**Corresponding Authors:** bruno-cesar.da-silva@neel.cnrs.fr and martien.den-hertog@neel.cnrs.fr

**Abstract**

Momentum resolved 4D-STEM, also called center of mass (CoM) analysis has been used to measure the long range built-in electric field of a Silicon p-n junction. The effect of different STEM modes and the trade-off between spatial resolution and electric field sensitivity are studied. Two acquisition modes are compared: nanobeam (NB) and low magnification (LM) modes. A thermal noise free Medipix3 direct electron detector with high speed acquisition has been used to study the influence of low electron beam current and millisecond dwell times on the measured electric field and standard deviation. It is shown that LM conditions can underestimate the electric field values due to a bigger probe size used, but provide an improvement of almost one order of magnitude on the signal-to-noise ratio (SNR), leading to a detection limit of $0.011 MVcm^{-1}$. It is observed that the CoM results do not vary with acquisition time or electron dose as low as 24 e⁻/A$^2$, showing that the electron beam does not influence the built-in electric field and that this method can be robust for studying beam sensitive materials, where a low dose is needed.

The continuous development of semiconductor devices demands techniques for assessing the electrostatic potential and related electric field created by doping contrast with high accuracy and spatial resolution. Imaging and quantifying built-in potential or electric fields at a p-n junction has been demonstrated by different STEM and TEM based techniques[1–6]. Among them, four-dimensional scanning transmission electron microscopy (4D-STEM) is a promising method because a complete diffraction pattern is recorded, allowing robust and quantitative measurements with unlimited field of view.

4D-STEM consists in scanning a thin sample with a focused electron beam and acquiring at each beam position **R** an electron diffraction pattern on a 2D-detector, with a diffraction intensity $I_D(\mathbf{k}, \mathbf{R})$, where k is a vector in reciprocal space, generating a 4D dataset. By processing this 4D-dataset, acquired with a convergent electron beam, the electric or magnetic fields in the specimen under study can be recovered. The principle of this electric field measurement is simply understood within the electron-particle model. Due to the Coulomb force, in the presence of a uniform electric field on a thickness $t$, an incoming electron beam is deviated from its initial direction by an angle $\gamma$. The electron deflection can be related to the modulus of the normal component of projected electric field along the sample thickness $t$ by:[4,7]

$$E_\perp^{proj} = -\frac{p_0^2}{m_e^* qt}\gamma \qquad (1)$$

where $m_e^*$ the relativistic mass of the electron, $q$ is the elementary charge, $p_0$ it is the initial electron momentum, and for a simple case $E_\perp^{proj}(r_\perp) = \frac{1}{t}\int E_\perp(x,y,z)dz$. On the other hand, using the wave nature of the electron beam and the phase object approximation, a more general relationship between the projected electric field and the beam deflection, described as a momentum transferred, can be obtained:[8,9]

$$\langle p_\perp(R)\rangle = -\frac{qt}{v} E_\perp^{proj}(r_\perp) * I_{probe}(r_\perp, R) \qquad (2)$$

where * denotes the convolution of the projected electric field $E_\perp^{proj}(r_\perp)$ with the probe intensity within the sample $I_{probe}(r_\perp, R)$. The momentum transferred and CoM performed in the diffraction pattern are connected by $\langle p_\perp(R)\rangle \equiv \frac{h}{2\pi} k_{COM}$, where $h$ is the Planck constant. This is the reason that the CoM-analysis is also called momentum resolved 4D-STEM. When the probe dimension is small with respect to the distance over which the electric field $E_\perp^{proj}(r_\perp)$ varies, the convolution in equation 2 can be neglected, resulting in the expression:[8]

$$\langle p_\perp(R)\rangle \equiv \frac{h}{2\pi} \iint \frac{k\, I_D(k,R)}{\mathbb{I}_D} dk = -\frac{qt}{v} E_\perp^{proj}(r_\perp) \qquad (3)$$

where $\mathbb{I}_D$ is the integrated diffraction pattern. Equations (1) and (3) are very similar, and show that by performing CoM in the diffraction pattern, the momentum transferred (or the associated beam deflection $\gamma$) can be measured and, consequently, the internal electric field recovered.

Typically, the built-in electric fields in semiconductor p-n junctions are small making 4D-STEM experiments challenging. For example, an abrupt Si p-n junction with doping levels of about $10^{17} - 10^{19} cm^{-3}$, should have an electric field in the range of $0.1 - 1.3 MV cm^{-1}$, and for the sample thickness used in STEM (100 – 300 nm) and 200 kV electrons, the beam deflections will be in the range of $3.0 - 100.0\ \mu rad$, which, depending on the experiment geometry, will correspond to a few pixels deflection or less. In 4D-STEM, as demonstrated in Differential Phase Contrast (DPC) performed with a four quadrant detector by Haas et al.[4], there is a trade-off between beam deflection sensitivity and spatial resolution[4,7] and it can result in different effects in the CBED patterns: rigid shift or redistribution of intensity[8,10].

In thicker samples, diffraction contrast changing across the sample can drastically affect the DPC and 4D-STEM imaging process[7,11]. To mitigate this challenge, the measurement is usually performed with the crystal in off-axis conditions, precessing the beam[7] or using a data analysis method based on template matching[7,12–15]. However, high resolution on axis conditions analyzed by CoM have been already used to measure the electric field at p-n junctions,[6] but the application to more challenging samples, for instance semiconductors with very small built-in electric field such as Silicon and Germanium, or graded junctions, is not obvious. Recently, Simon Pöllath et al.[9] have studied by theoretical considerations and simulations the effect of illumination conditions on the precision of the calculated CoM in pixelated detectors.

So far, no experimental work has investigated if momentum resolved 4D-STEM can provide feasible and reliable results in challenging systems with very small built-in electric fields when performed in different acquisition conditions, especially when a low electron dose is used in order to avoid modification in the intrinsic electrical properties of the sample. In this work, the effect of different

STEM modes, as well as the influence of electron beam current and dwell time on the measured electric field and corresponding standard deviation is studied on a silicon p-n junction, measured by 4D-STEM applying CoM and using a thermal noise free direct electron detector with high frame rates.

The sample is composed of a symmetrically doped n-type phosphorus and p-type boron silicon p-n junction with $9 \times 10^{18} cm^{-3}$ doping concentration. The specimen was prepared using a FEI Strata dual beam focused ion beam (FIB) and optimized protocols in order to reduce the gallium implantation[1]. The p-n junction silicon lamella thickness was 350 nm and a 60 nm dead layer for the same sample has been reported[16]. The specimen was mounted for STEM observation in a Protochips Aduro 500 sample holder without applying any bias to the contacts.

The 4D-STEM experiments were carried out on a FEI Titan Ultimate aberration-corrected (S)TEM microscope operated at 200 kV equipped with a fast pixelated (256 × 256 pixels) Medipix 3 based Merlin camera. Two settings have been used for the measurements: i) a nano-beam (NB-STEM) mode, with nominal semi-convergence angle of 1.1 mrad and a camera length of 2.3 m and ii) a low magnification STEM mode (LM-STEM), using a nominal semi-convergence angle of 0.18 mrad and a camera length of 13.5 m. In both settings there were no overlapping diffracted disks, and the transmitted beam covered the central part of the detector. The angular pixel size in the pixelated detector was 11 μrad in NB and 2.2 μrad in LM settings, respectively. A pixel size in real space of 5 nm was used in NB, while 6 nm was employed in LM settings. Different diffraction maps were acquired using dwell times in the range of 1 – 100 ms, which took between 2 - 33 min. A beam current of 30 pA and 100 pA was used in different sets of experiments. The sample was tilted off axis in an orientation where the diffraction contrast was minimized.

Data analyses were performed using python scripts and the CoM of each CBED pattern for each beam position was computed. Typical CBED patterns in NB and LM are shown in Figure 1d and 1g, respectively. No additional treatment such as masks or filters were applied during data analysis. Using a reference region outside the p-n junction, the CoM beam deflection was estimated and converted to electric field using equation 1 [2,17]. The atomic electric fields were removed by employing a probe size larger than the unit cell of the material under study in both NB and LM modes.

Working in Nano-Beam (NB-STEM) settings, Figure 1c, smaller probe sizes can be obtained by using a larger semi-convergence angle of 1.1 mrad and, consequently, better spatial resolution in real space. However, a short camera length (CL 2.3 m) has to be used in order to physically fit the transmitted beam in the detector dimensions. Figure 1d and 1e present the CBED disks, comparing the disk profiles of the electron beam either traversing a field free sample region (blue) or traversing the p-n junction region with maximum electric field (red). Interestingly, the schematic view shown in Figure 1a and 1b works well here, and a probe smaller than the region where the electric field varies results in a so-called hard shift of the disk, as presented in Figure 1b. Thus, in this case, the effect of the electric field on the incoming beam can be easily interpreted by the electron-particle approach of an electron beam being deflected by the electric field due to the Lorentz force.

On the other hand, the use of smaller semi-convergence angle and longer camera length increases the beam deflection sensitivity, since a more magnified CBED is obtained, Figure 1f. However, Simon Pöllath et al. have suggested that the precision depends on the semi-convergence angle used, and not necessarily on the camera length[9]. In this setting, a degradation of spatial resolution is expected, since the probe size is typically increased due to the smaller beam convergence angle. A semi-convergence angle of 0.18 mrad and 13.5 m camera length were used by turning off the objective lens, in a mode known as LM. In Figures 1g and 1h it can be seen that in this setting the resulting CBED pattern presents a clear response with the electric field, but this effect is a complex redistribution of intensity. Therefore, if the data is acquired in LM mode, the use of template matching algorithms to detect the beam deflection would not make much sense, while CoM can capture the effect of the built-in electric fields in both settings.

In Figure 2a the results of the electric field maps measured in NB and LM by 4D-STEM are shown, analyzed by CoM. A clear signal is obtained at the p-n junction, without any artefact, but it can be observed that there is a fluctuation in the electric field values measured, which is greater in NB than in LM mode, as we can see in Figure 2a. These variations observed in the measurements reflect the precision related to the experimental settings, NB or LM, as well as the number of counts in the CBED pattern and intrinsic uncertainties related to the measurement being for example sample drift, thickness variations etc. The average value of the electric field in NB was $E_{max} = (0.16 \pm 0.09)\ MVcm^{-1}$, resulting from a beam deflection of $\gamma_{shift} = (13.5 \pm 7.6)\ \mu rad$. In LM, a lower field value of $E_{max} = (0.112 \pm 0.011)\ MVcm^{-1}$, which corresponds to a beam deflection of $\gamma_{shift} = (9.45 \pm 0.94)\ \mu rad$, was measured with a better precision. The values measured in NB are in good agreement with the result of off-axis electron holography performed in the same wafer[4], suggesting that the values in LM are underestimated. Further analysis and quantification of the measured electric field is beyond the scope of the present article and will be published elsewhere.

Figure 2b-c presents line profiles over the electric field maps shown in Figure 2a in a region where the electric field value corresponds to the average electric field. It is possible to see that NB mode provides a poor SNR compared to LM mode. The SNR of the field profiles in NB conditions can be improved by performing an integration over a wider region in the electric field map, providing better SNR profiles across the junction, Figure 2c. However, this approach will present some limitations in junctions restricted to specific regions, like in nanowires for example. For the studied p-n junction sample, a depletion region in the range of (70 - 100) nm is observed in both 4D-STEM modes, Figure 2b and 2c. A summary of the parameters and values found in the measurements performed in NB and LM mode is shown in Table 1. We can see that the decrease of semi-convergence angle by a factor of six has enhanced the SNR by the same amount in LM, compared to NB[9].

In table 1 we can see that the LM result appears to be approximately 30% reduced compared to the NB results. This is likely due to the bigger probe size in LM. We have measured a 3.6 nm (FWHM) probe size in NB conditions. The probe size in LM is more challenging to measure precisely due to the low magnification available, but according to Abbe's law, a probe size of around 10 nm in LM conditions is expected. In Figure 2d it is possible to see qualitatively that the convolution of the expected electric field, as obtained in NB settings (which we assume to be due to a graded p-n junction)[18] with different probe sizes, leads to a reduction and broadening of the measured electric field for increasing probe sizes. Figure 2d offers a qualitative understanding, since a quantitative analysis should take the broadening, propagation and absorption of the probe within the interaction volume in the sample into account.

In order to better understand how the other acquisition conditions, affect the SNR of the electric field maps, the standard deviation was measured at different experimental conditions. Figure 3a shows the dependence of $3\sigma_E$ with the average counts per pixel in the pixelated detector. The counts were modified either by changing the beam current or the pixel dwell time. It is possible to see that the decrease in the uncertainty with the increase in counts quickly saturates, and no substantial improvement is obtained in the SNR for acquiring very well defined CBED patterns, Figure 3a. The standard deviation was extracted in the region with no field. The small fluctuations observed in Figure 3a are likely due to some remaining influence of diffraction contrast or sample thickness variations, but these variations are small and the general trend can be clearly observed, as predicted by simulations[9]. The smallest standard deviation measured results in $3\sigma_E = 0.074\ MVcm^{-1}$ in NB mode, and $3\sigma_E = 0.011\ MVcm^{-1}$ in LM mode, representing the detection limits in both NB and LM mode on our sample. The values shown in Figure 3a represent a real achievable uncertainty, since the dynamical effect will be always present even when they are minimized by acquiring maps in off-axis orientation.

The results of these different acquisition conditions in the quantification of the electric field are shown in Figure 3b-f in LM mode. Well-defined CBED patterns can be obtained using long dwell times, Figure 3d, but the electric field maps and profiles provide almost identical results, Figure 3f. The small difference between the maximum electric field measured is below the uncertainty of the measurement, as shown in Figure 3a. The same behavior was observed for the NB measurements. Thus, our results show that low dose CoM analysis is a promising method for beam sensitive materials, as quantitative and reproducible results can be obtained using CoM even at an electron dose as low as 24 e⁻/Å².

The differences between LM and NB settings seems to be not related to the pixel size in reciprocal space. Figure 4a-f and Table S1 show the effect of pixel binning of the CBED patterns on the electric field measurements: the increase in pixel size due to binning has no effect on either the signal-to-noise ratio (SNR), or the electric field measured. In Figure 4g and 4h the effect of both semi-convergence angle as well as camera length on $\sigma_E$ is investigated. Since the electric field standard deviation also depends on the number of counts we have marked the average counts per pixel of each measurement for better comparison. Figure 4g shows that for decreasing semi-convergence angle (keeping the camera length constant) the noise level is reduced. The same behavior is observed by increasing the camera length (at constant semi-convergence angle), Figure 4h. This result is in agreement with the proposal of Pöllath et al.[9], who have suggested that the electric field uncertainty is intrinsically linked to the semi-convergence angle, due to the Heisenberg's uncertainty relation. Nevertheless, our results show that the camera length also plays a role.

The present study benefitted from the Merlin 1R detector, allowing milliseconds pixel times with excellent detector performance, see comments in supporting information S2. This is a clear advantage, since more traditional detectors often need about an order of magnitude longer pixel time at their fastest acquisition rate, making the experiment more prone to sample drift and damage.

Although the electric field measured in LM appears to be underestimated, LM provides much better SNR maps, which are better suited to localize very small long-range electric fields. The methodology presented here may also be suitable for studying more advanced semiconductor devices as FinFETs or gate all around devices, composed of homojunctions. The key points are the proper sample preparation and the adjustment of the settings to achieve a better SNR without losing the spatial resolution needed. Although laborious, the sample preparation by focused ion beam of 3D devices for field mapping using TEM based techniques is already a reality.[19] The projection of the 3D electric field can be studied by producing a lamella extending parallel to the gate length (the distance between the source and the drain) and perpendicular to the fin width. The analysis will be limited by the thickness of the cross sectional lamella: the fin width. The gate length will determine the spatial resolution needed. For example, for a sample thickness of 30 nm (fin width), an electric field of $E_{max} = 0.12\ MV cm^{-1}$ and 200 keV electrons, a beam deflection of around 1 µrad is expected, approaching the best detection limit demonstrated here ($3\sigma_{shift} \cong 0.9$ µrad ). The optimal settings will therefore depend on the sample and a comprise between the SNR and the spatial resolution should be found, taking into account the strength and spatial distribution of the expected built-in electric field.

In summary, a silicon p-n junction lamella has been examined by 4D-STEM using CoM analysis and different acquisition conditions: nanobeam (NB) and low magnification (LM) mode. In NB, a hard shift of the transmitted disk was observed, while in LM conditions a redistribution of intensity in the CBED pattern occurs. In LM, an underestimation of electric field maximum compared to NB was observed, likely related to the larger electron probe size. However, a better SNR is achieved using a smaller semi-convergence angle and longer camera length available in LM mode. It is shown that a detection limit as good as $0.011 MV cm^{-1}$ could be obtained in LM and the results of the CoM analysis on the CBED patterns remained reproducible even at very low electron dose at or below 24 e$^-$/Å$^2$.

These results evidence that small, long range built-in electric fields at homojunction p-n samples present in beam sensitive materials can be studied and localized by 4D-STEM and CoM at low electron dose, in conditions where the electric field sensitivity is enhanced, while maintaining a spatial resolution in the nm range.

**Supplementary Material**

See supplementary material for additional information about the pixel binning effect on the electric field measurement and a brief description of the advantages and important characteristics of the detector used.

**Acknowledgements**

This project received funding from the European Research Council under the European Union's H2020 Research and Innovation programme via the e-See project (Grant No. 758385). These experiments have been performed at the Nanocharacterisation platform PFNC in Minatec, Grenoble.

**Data Availability Statement**

The data that support the findings of this study are available from the corresponding authors upon reasonable request.

**Notes**

The authors declare no competing financial interest.

**TABLE**

**Table 1** - Comparison of LM-STEM vs. NB-STEM for the performed 4D-STEM experiments realized in a Si p-n junction, corresponding to the data shown in Fig. 2. The signal-to-noise ratio is calculated by dividing the total signal by the noise, $3\sigma$. The standard deviation was calculated from the region of no field (area of 70 nm x 60 nm).

| STEM Mode | **LM-STEM** | **NB-STEM** |
|---|---|---|
| Settings | | |
| Dwell time (ms) | 5 | 5 |
| Beam current (pA) | 100 | 100 |
| Semi-convergence Angle (mrad) | 0.18 | 1.1 |
| Camera Length (m) | 13.5 | 2.3 |
| Pixel size real space (nm) | 6 | 5 |
| Pixel size reciprocal space (µrad) | 2.2 | 11.0 |
| $\overline{N}_{\text{per pixel}}^{\text{CBED}}$ (counts) | 10.5 | 14.5 |
| **Measurements – Beam Deflection** | | |
| Beam deflection $\delta_{\max}^{\text{measured}}$ (pixels) | 4.30 | 1.23 |
| Noise level $3\sigma_\delta$ (pixels) | $3\sigma_\delta^{\text{LM}} = 0.41$ | $3\sigma_\delta^{\text{NB}} = 0.61$ |
| **Measurements – Electric Field** | | |
| $E_{\max}^{\text{measured}}$ (MVcm$^{-1}$) | 0.112 | 0.160 |
| Noise level $3\sigma_E$ (MVcm$^{-1}$) | $3\sigma_E^{\text{LM}} = 0.011$ | $3\sigma_E^{\text{LM}} = 0.087$ |
| Signal to noise ratio (SNR) | 10.2 | 1.8 |

# FIGURE CAPTIONS

**Figure 1** – Schematic of the Electric field measurement through the detection of the beam deflection on a pixelated detector. (a) The electric field within the p-n sample is probed by scanning the sample with a convergent electron beam. (b) the CBED transmitted disk is acquired at each scan position. The blue and red circles represent the direct beam of the diffraction pattern at different locations in the sample: at a reference point without (blue) and with and electric field present in the sample (red). Measuring the shift of the direct beam caused by the built-in electric field, the magnitude and direction of electric field can be measured. (c, f) Schematics showing the differences in the measurement's conditions for NB 4D-STEM (c, d, e) and LM 4D-STEM (f, g, h), respectively. (d, g) Example of raw data: CBED patterns acquired in the sample region where there is no field (blue) and in the center of the p-n junction (red). (e, h) Profiles across these CBED patterns

**Figure 2** – (a) Electric field maps of a silicon p-n junction acquired in NB-STEM and LM-STEM modes, using a beam current of 100 pA and 5 ms dwell time, (b, c) profiles of the maps in (a). (d) Black line: Expected electric field profile of a p-n junction with $E_{max} = 0.16\ MVcm^{-1}$ and $W_d = 70\ nm$. Colored lines: convolution of different probe sizes in the range of 2 to 40 nm with the electric field profile in black.

**Figure 3** – (a) Dependence of the noise level of the measurement of the electric field, $3\sigma_E$, on the number of counts and the measurement conditions. The standard deviation was calculated from the region of no field (area of 70 nm x 60 nm). CBED patterns acquired with different dwell times (b) 1 ms, (c) 10 ms and (d) 100ms, with a beam current of 100 pA. The uncertainty in the electric field corresponding to the measurement conditions of each CBED pattern, marked I (a), II (b) and III (c), is highlighted in (a) with the corresponding number at the data point. (e) 4D-STEM maps for different dwell times. (f) Electric field profiles extracted from (e) at the indicated location integrated over 100 nm.

**Figure 4** – (a) to (d) Example of CBED pattern acquired in LM conditions with different pixel binning, using a semi-convergence angle of 0.20 mrad, camera length of 13.5 m, dwell time of 10 ms and pixel size of 6 nm in real space. (a) 256 x 256 pixels, not binned. (b) 128 x 128 pixels. (c) 64 x 64 pixels. (d) 32 x 32 pixels. (e) Electric field maps and (f) corresponding profiles obtained using the data shown in (a) - (d). (g) and (h) Dependence of the noise level of the electric field, $3\sigma_E$, on different measurement conditions. (g) Different semi-convergence angles, keeping the camera length constant at 13.5 m. (h) Different camera lengths, keeping the semi convergence angle α at 0.25 mrad. The average counts per pixel of each measurement is indicated in the legend.

**Figure 1**

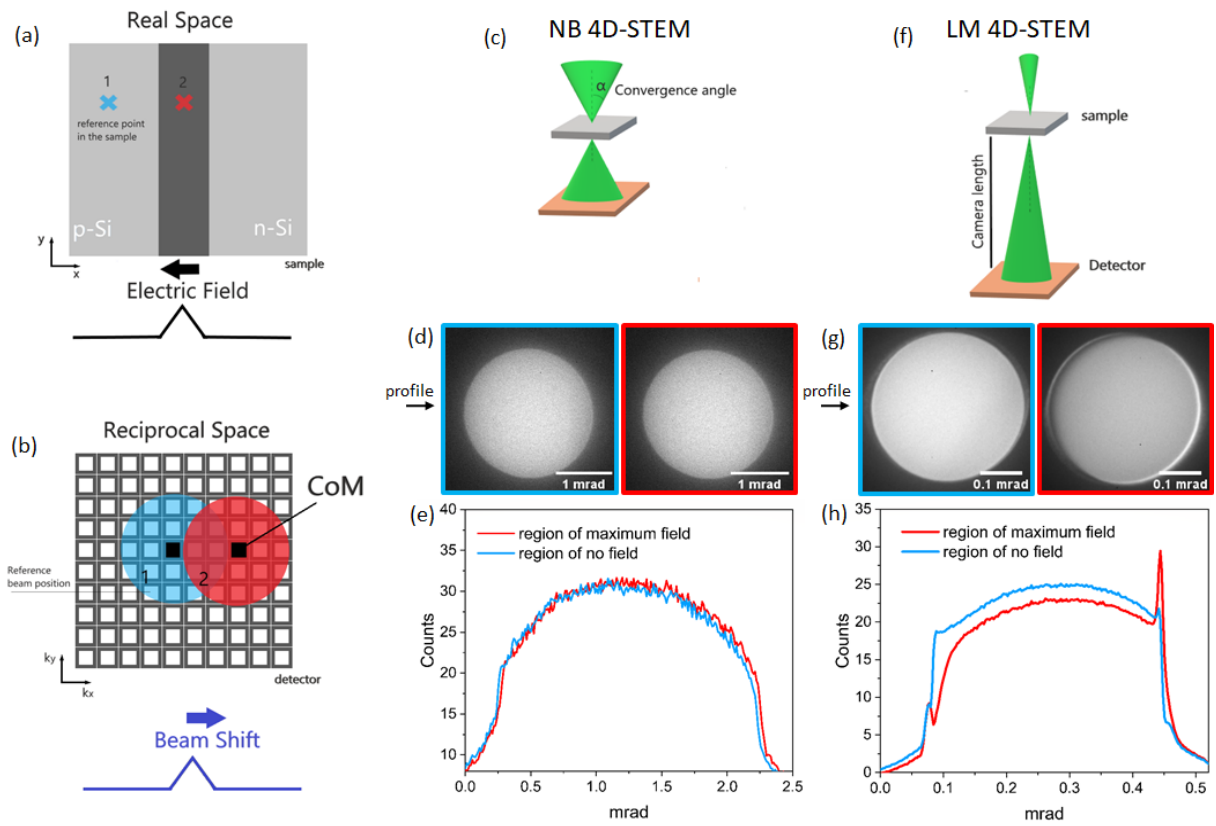

**Figure 2**

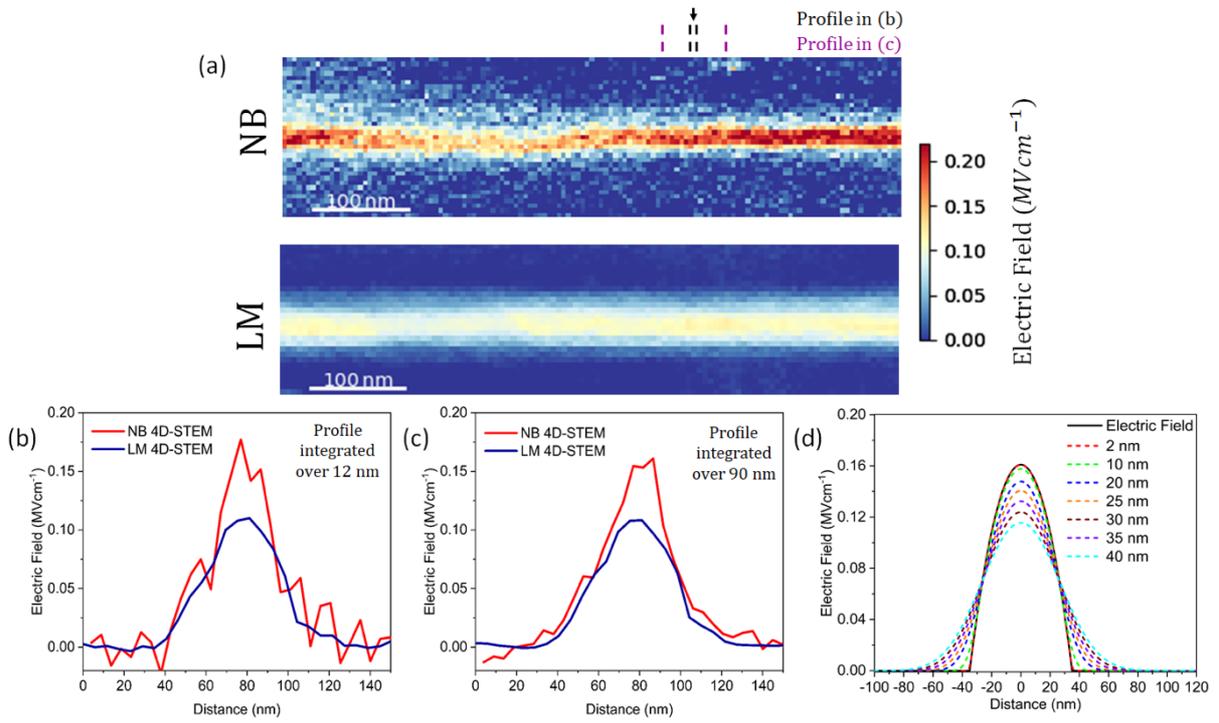

**Figure 3**

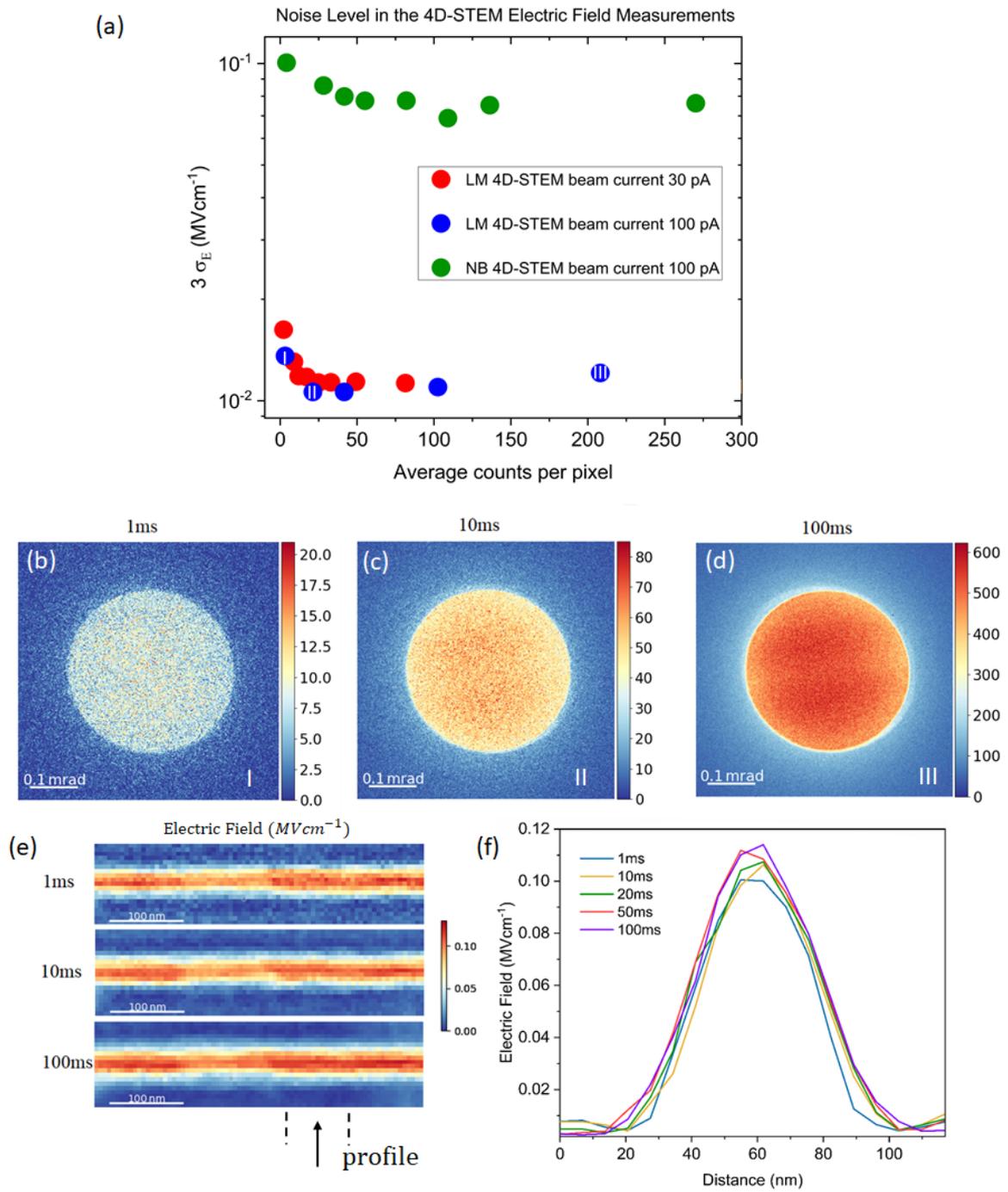

**Figure 4**

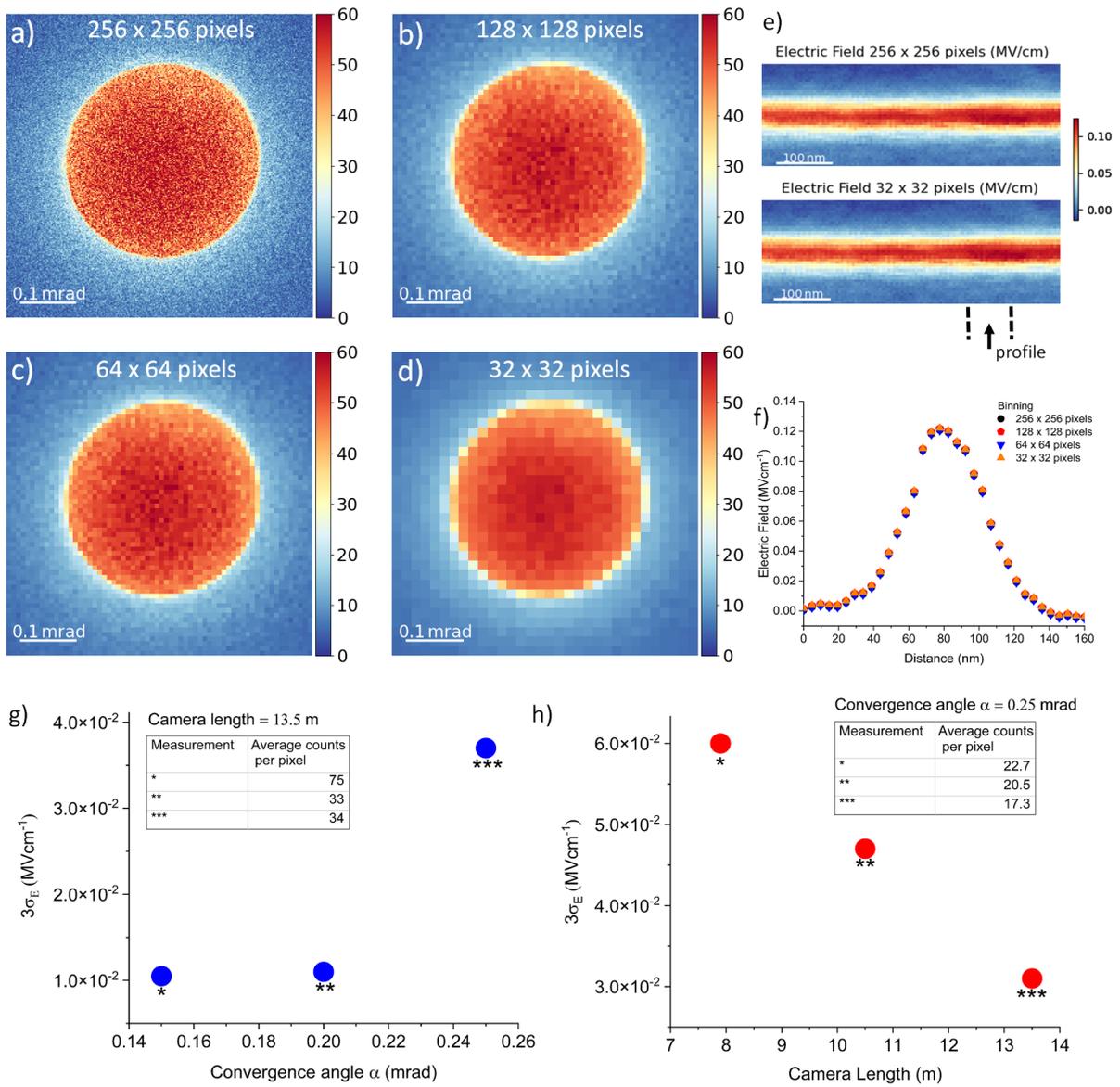